\documentclass[twocolumn,english,showpacs,pra]{revtex4}
\usepackage[T1]{fontenc}
\usepackage[latin9]{inputenc}
\usepackage{textcomp}
\usepackage{amsmath}
\usepackage{graphicx}
\usepackage{amssymb}

\makeatletter

\providecommand{\tabularnewline}{\\}

\@ifundefined{textcolor}{}
{%
 \definecolor{BLACK}{gray}{0}
 \definecolor{WHITE}{gray}{1}
 \definecolor{RED}{rgb}{1,0,0}
 \definecolor{GREEN}{rgb}{0,1,0}
 \definecolor{BLUE}{rgb}{0,0,1}
 \definecolor{CYAN}{cmyk}{1,0,0,0}
 \definecolor{MAGENTA}{cmyk}{0,1,0,0}
 \definecolor{YELLOW}{cmyk}{0,0,1,0}
 }

\usepackage{dcolumn}\usepackage{bm}\usepackage{amsfonts}\usepackage{times}\usepackage{babel}\usepackage{babel}

\makeatletter
\@ifundefined{textcolor}{}
{
 \definecolor{BLACK}{gray}{0}
 \definecolor{WHITE}{gray}{1}
 \definecolor{RED}{rgb}{1,0,0}
 \definecolor{GREEN}{rgb}{0,1,0}
 \definecolor{BLUE}{rgb}{0,0,1}
 \definecolor{CYAN}{cmyk}{1,0,0,0}
 \definecolor{MAGENTA}{cmyk}{0,1,0,0}
 \definecolor{YELLOW}{cmyk}{0,0,1,0}
 }
\providecommand{\U}[1]{\protect\rule{.1in}{.1in}}
\makeatother
\makeatother

\makeatother

\usepackage{babel}

\makeatother

\usepackage{babel}

\makeatother

\usepackage{babel}

\makeatother

\usepackage{babel}

\begin{document}

\title{Magic doping: From the localized hole-pair to the checkerboard patterns}

\author{X. Q. Huang$^{1,2}$}

\email{xqhuang@netra.nju.edu.cn}

\affiliation{$^{1}$Department of Physics and National Laboratory of Solid State
Microstructure, Nanjing University, Nanjing 210093, China\\
 $^{2}$Department of Telecommunications Engineering ICE, PLAUST,
Nanjing 210016, China }

\date{\today}
\begin{abstract}
Intensive experiments have revealed that the superconductivity of
the hole-doped cuprates can be strongly suppressed at the so-called
magic doping fractions. Despite great research efforts, the origin
of the `magic doping' remains mysterious. Recently, we have developed
a real-space theory of high-temperature superconductivity which reveals
the intrinsic relationship between the localized Cooper pair and the
localized hole pair (arXiv:1007.3536). Here we report that the theory
can naturally explain the emergence of non-superconducting checkerboard
phases and the magic doping problem in hole-doped cuprate superconductors.
It clearly shows that there exist only seven `magic numbers' in the
cuprate family at $x=$1/18, 1/16, 2/25, 1/9, 1/8, 2/9 and 1/4 with
$6a\times6a$, $4a\times4a$, $5a\times5a$, $3a\times3a$, $4a\times4a$,
$3a\times3a$, and $2a\times2a$ checkerboard patterns, respectively.
Moreover, our framework leads directly to a satisfactory explanation
of the most recent discovery {[}M. J. Lawler, \emph{et al}. Nature
\textbf{466}, 347 (2010){]} of the symmetries broken within each copper-oxide
unit in hole-doped cuprate superconductors. These findings may shed
new light on the mechanism of superconductivity.
\end{abstract}


\maketitle
The `1/8 anomaly' \cite{Moodenbaugh1988}, one of the long-standing
puzzles of superconducting physics, is believed to be a key to understanding
the mechanism of high-$T_{c}$ superconductivity. Later, a number
of experiments have indicated that the anomalous suppression of superconductivity
can be observed in the hole-doped ($p$-type) cuprates at other `magic'
hole densities, for example, 1/16 \cite{Zhou2004,Zhou2003} and 1/9
\cite{Zhou2004}. Great efforts have been made to determine these
`magic numbers'. The SO(5) theory predicts a series of magic doping
fractions at $x=(2m+1)/2^{n}$, where $m$ and $n$ are integers \cite{Chen2004},
while Feng $\textit{et al.}$ \cite{Feng2006} obtained a single-parameter
expression as $x=(2+n^{2}-4n)/n^{2}$, where $n=4,5,6,\cdots$. These
interpretations imply the possibility of an infinite magic doping
fractions in $p$-type cuprate superconductors, which is obviously
inconsistent with the experimental facts. In addition, the physical
meanings of the integers $n$ and $m$ are not very clear in these
results. In our opinion, these theoretical results cannot be expected
to be physically correct and reasonable.

Recently, we have proposed a universal mechanism for the superconductivity
which offers a new way of looking at the superconducting phenomenon
\cite{Huang1,Huang2}. In particular, we have introduced a new model
for $d$-wave pairing in hole-doped high-$T_{c}$ superconductors
which is able for the first time to satisfactorily describe the pseudogap
related phenomena, such as the Fermi pocket (or Fermi arc), the two
pseudogap behavior and the linear relationship between the pseudogap
temperature $T^{\ast}$ and the hole doping level $x$ in these compounds
\cite{Huang3}.

\begin{figure}[b]

\begin{centering}
\resizebox{1\columnwidth}{!}{ \includegraphics{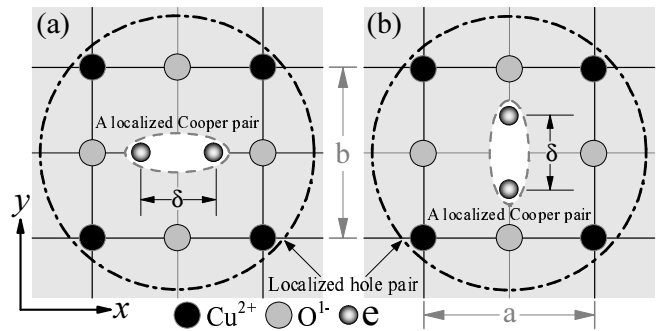}}
\par\end{centering}

\caption{The relationship between the localized Cooper pair and the localized
hole pair in the hole-doped cuprates. (a) Two electrons arranged along
the $x$-direction, (b) two electrons aligned in the $y$-direction.
If $a=b$, we have proved analytically that the repulsive interactions
among Cooper pairs have been completely suppressed with the appropriate
$\delta\approx0.396b$ \cite{Huang3}.}

\label{fig1}
\end{figure}

In the present paper, based on the new theory and model, we aim to
uncover the underlying relationship between the `magic numbers' and
the nature charge ordering of `checkerboard' as suggested by neutron
and STM experiments \cite{Hoffman2002,Hanaguri2004,McElroy2005}.
Seven magic numbers and the corresponding checkerboard structures
are analytically and uniquely determined for the $p$-type cuprates.
One will easily find that our results are completely different from
those obtained through quantum theory and quantum field theory \cite{Chen2004,Feng2006}.
Finally, we will show that the suggested model can explain the recent
experiments of the rotational symmetry breaking of the pseudogap phases
in hole-doped cuprate superconductors \cite{Daou2010,Lawler2010}.

\section{Localized Cooper pair and hole pair}

The hole-doped cuprates have been intensively investigated because
of the relatively high superconducting transition temperature and
a rich phase diagram. However, the question of what causes the loss
of electrical resistance in these materials is still one of the major
unsolved problems in physics. We have argued that the main reason
for this situation is that researchers are confused about some fundamental
ideas of modern physics, for example, what is the `hole'? As emphasized
by Hirsch \cite{Hirsch2005}, using the language of `holes' rather
than `electrons' in fact obscures the essential physics since these
electrons are the ones that `undress' and carry the supercurrent (as
electrons, not as holes) in the superconducting state.

In our scenario model \cite{Huang3}, a hole is a real-space `quasiparticle'
which is composed of some well-known electrons and ions. For the hole-doped
cuprates, a localized hole-pair is a cluster of two electrons (a localized
Cooper pair), four $O^{1-}$ and four $Cu^{2+}$ inside the Cu-O plane,
as illustrated in Fig. \ref{fig1}. According to the classical electromagnetic
theory, it is very easy to prove that the direct and strong electron-electron
repulsion can be entirely excluded if two electrons are aligned in
$x$-direction {[}Fig. \ref{fig1}(a){]} or $y$-direction {[}Fig.
\ref{fig1}(b){]} within each copper-oxide unit with the Cooper-pair
size $\delta\approx0.396b$ (when $a=b$). More importantly, we have
shown analytically and numerically that the nearest-neighbor electron-$O^{1-}$
repulsive interactions play the key role of the `pairing glue' for
the real-space localized Cooper pair. The simply picture of Fig. \ref{fig1}
could yield a pairing and superconducting scenario that has the potential
to resolve the pseudogap and the high-$T_{c}$ superconducting puzzles
in the hole-doped cuprates. In our series of studies, we will show
that the `'hole' pictures (Fig. \ref{fig1}) are considered to be
the most basic structural units of the hole-doped cuprates, which
can further self-assemble into some superconducting electron states
and non-superconducting pseudogap and checkerboard phases. All the
related physical properties, such as vortex lattices, Meissner effect,
London penetration depth, Hall effect, $d$-wave symmetry, checkerboard
patterns, magic doping fractions, Fermi pocket (or Fermi arc), two
pseudogap behavior and rotational symmetry breaking, can be perfectly
interpreted by our framework.

\section{Non-superconducting checkerboard patterns}

\begin{figure}
\begin{centering}
\resizebox{1\columnwidth}{!}{ \includegraphics{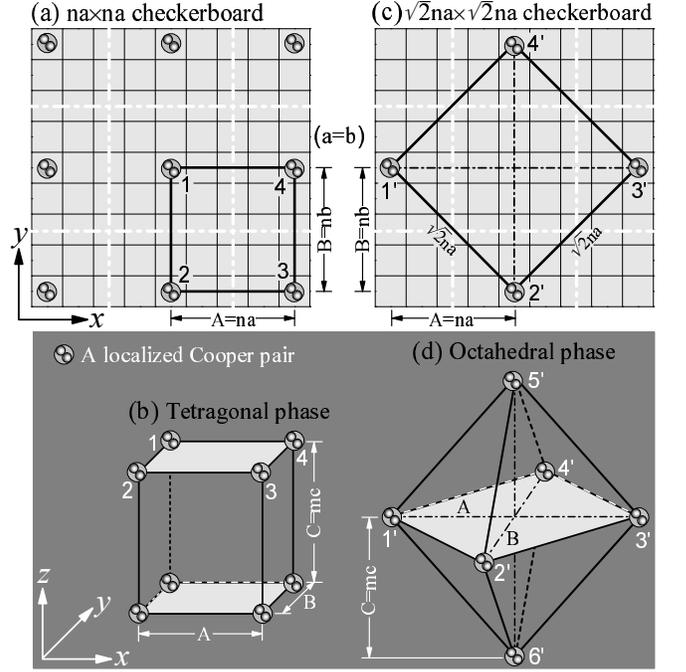}}
\par\end{centering}

\caption{Two non-superconducting vortex lattices.\textbf{ }(a) and (b) The
tetragonal phase, where the localized Cooper pairs can form a $na\times na$
checkerboard pattern in the charge-carrier doped planes. (c) and (d)
The octahedral phase, where the $\sqrt{2}na\times\sqrt{2}na$ checkerboard
can be found in each doped plane. }

\label{fig2}
\end{figure}

The question remains open as to why is the superconductivity suppressed
in the cuprates at the magic doping levels? In our opinion, the suppression
of superconductivity is caused by a electronic structure phase transition
from a superconducting state to a localized state. It is most likely
that the charge carries are pinned via Coulomb interaction with their
associated ions. From the perspective of symmetry breaking, the electrons
in the localized state must have a higher symmetry than that of the
superconducting state. Two high symmetry non-superconducting Wigner
crystals of localized Cooper pair of Fig. \ref{fig1} are suggested
and shown in Fig. \ref{fig2}.

Figs. \ref{fig2}(a) and (b) represent the tetragonal phase, where
the $na\times na$ checkerboard pattern can form in all doped planes
{[}see Fig. \ref{fig3}(a){]}. If one of $m$ planes in the layered
superconductor are doped, thus the corresponding doping level is given
by \begin{equation}
x=p_{1}(n,n,m)=2\times\frac{1}{n}\times\frac{1}{n}\times\frac{1}{m}=\frac{2}{n^{2}m}.\label{x_NSC1}\end{equation}

For the case of the octahedral phase of Figs. \ref{fig2}(c) and (d),
the checkerboard shows $\sqrt{2}na\times\sqrt{2}na$ structure in
each doped plane {[}see Fig. \ref{fig2}(c){]}. It is easy to obtain
the doping level for this phase as follows \begin{equation}
x=p_{2}(n,n,m)=2\times\frac{1}{\sqrt{2}n}\times\frac{1}{\sqrt{2}n}\times\frac{1}{m}=\frac{1}{n^{2}m}.\label{x_NSC2}\end{equation}

Evidently, such commensurate stripes are unmovable and should be insulating.
Note from Fig. \ref{fig2} that when $C=A=B$, the tetragonal and
octahedral phases will transit into the more stable cubic and regular
octahedral phases, respectively. In the following section we will
show that the perfect cubic and regular octahedral phases can be found
in $Ca_{2-x}Na_{x}CuO_{2}Cl_{2}$ at $x=$1/8 and 1/16, respectively.

\section{Checkerboard and magic doping }

\begin{figure}
\begin{centering}
\resizebox{1\columnwidth}{!}{ \includegraphics{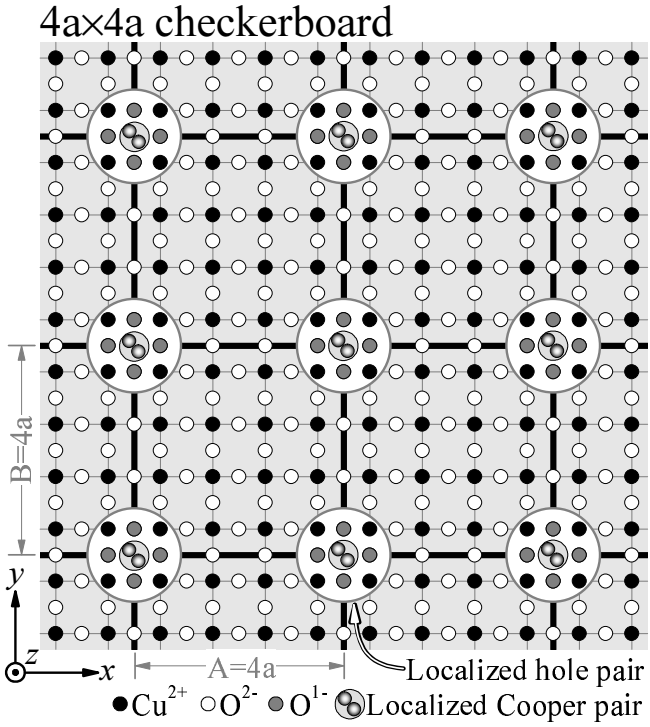}}
\par\end{centering}

\caption{The $4a\times4a$ checkerboard in hole-doped $CuO_{2}$ plane at doping
level $x=$1/8.\textbf{ }A pure electron picture of the localized
hole pairs, where the localized hole pairs of Fig. \ref{fig1} can
selforganize into a periodic nondispersive superlattices due to the
real space Coulomb confinement effect. }

\label{fig3}
\end{figure}

Fig. \ref{fig3} shows the purely electronic description of the $4a\times4a$
checkerboard in hole-doped $CuO_{2}$ plane, where the structural
relationship between the localized hole pair (the geometry of a cluster
of two electrons, four $O^{1-}$ and four $Cu^{2+}$) and the localized
Cooper pair is clearly illustrated. In a previous paper \cite{Huang3},
we proved analytically that the localized Cooper pair is a pure nearest-neighbor
confinement effect of the four nearest-neighbor $O^{1-}$, as indicated
in Fig. \ref{fig1}. Moreover, it has been evidently shown that a
localized Cooper pair is stable only when the two electrons aligned
in $x$- or $y$-directions, supporting the $d$-wave pairing symmetry.
The nearest-neighbor character of the pairing mechanism implies that
the localized Cooper pairs (pseudogap) are likely to survive in insulating
or nonmetallic materials, as was reported recently by Stewart $\textit{et al.}$
\cite{Stewart2007}.

It is now well accepted that the 1/8 anomaly is always accompanied
by the appearance of the $4a\times4a$ checkerboard in the hole-doped
cuprates \cite{Hoffman2002}. In fact, equations (\ref{x_NSC1}) and
(\ref{x_NSC2}) imply the existence of some sort of connection between
the magic doping fractions and the checkerboard patterns. According
to equation (\ref{x_NSC1}), when all $CuO_{2}$ planes are doped
$(m=1$), the $4a\times4a$ checkerboard pattern of Fig. \ref{fig3}
may be obtained in the corresponding cuprate sample at \textbf{$x=2/4^{2}=1/8$}.
As it is well known that the cubic phase is more stable than the tetragonal
phase, thus we define the following ratio of the lattice constants
$\delta=|\left(C-A\right)/C|=|\left(mc-na\right)/mc|$ which can be
used to estimate qualitatively the stability of the checkerboard phase
of different superconductors. For the three typical cuprates {[}$YBa_{2}Cu_{3}O_{y}$
($YBCO$: $a=3.87\textrm{\AA}$ and $c=11.72\textrm{\AA}$), $La_{2-x}Sr_{x}CuO_{4}$
($LSCO$: $a=3.79\textrm{\AA}$ and $c=13.25\textrm{\AA}$) and $Ca_{2-x}Na_{x}CuO_{2}Cl_{2}$
($CNCOC$: $a=3.80\textrm{\AA}$ and $c=15.18\textrm{\AA}$){]} with
the 1/8 anomaly, it is easy to get $\delta_{YBCO}\simeq0.321$, $\delta_{LSCO}\simeq0.144$
and $\delta_{CNCOC}\simeq0.001$. Obviously, these results indicate
that a `perfect' cubic structure of the localized Cooper pairs can
naturally form in $CNCOC$ (but not in $YBCO$ and $LSCO$) at $x=1/8$,
as a result, the $4a\times4a$ checkerboard phase in $CNCOC$ is much
more stable than that in $YBCO$ and$LSCO$ samples. Experimentally,
the $4a\times4a$ checkerboard pattern is better detected in $CNCOC$
superconductor \cite{Hanaguri2004}.

\begin{table}[tbp]
\caption{The possible magic doping fractions $\mathit{x}$ for $p$-type
cuprates when all the Cu-O planes are doped ($m=1$). }
\label{table1}%
\begin{ruledtabular}
\begin{tabular}{c|cccccc|ccccc}
Phase  & \multicolumn{6}{c|}{Tetragonal phase} & \multicolumn{5}{c}{Octahedral phase}\tabularnewline
\hline
$n$  & 2  & 3{*}  & 4{*}  & 5{*}  & 6{*}  & 7  & 1  & 2{*}  & 3{*}  & 4{*}  & 5\\
$x$  & $\displaystyle
{\frac{1}{2}}$  & $\displaystyle
{\frac{2}{9}}$  & $\displaystyle
{\frac{1}{8}}$  & $\displaystyle
{\frac{2}{25}}$  & $\displaystyle
{\frac{1}{18}}$  & $\displaystyle
{\frac{2}{49}}$  & 1  & $\displaystyle
{\frac{1}{4}}$  & $\displaystyle
{\frac{1}{9}}$  & $\displaystyle
{\frac{1}{16}}$  & $\displaystyle
{\frac{1}{25}}$\\
\end{tabular}
\end{ruledtabular}
\end{table}

It is immediately seen that in our framework the magic doping fractions
are closely correlated with the local checkerboard patterns. Based
on equations (\ref{x_NSC1}) and (\ref{x_NSC2}), one can obtain all
the relevant magic doping fractions for hole-doped cuprate superconductors
with the full-doped $CuO_{2}$ planes ($m=1$), as shown in Table
\ref{table1}. As we all know, the $T_{c}$ of hole-doped cuprates
has a dome-like shape as a function of hole concentration ranged from
$x\approx0.05$ to $0.27$, under this restriction, only seven `magic
numbers' ($x=$1/18, 1/16, 2/25, 1/9, 1/8, 2/9 and 1/4) are possible
in the cuprate family. Clearly, our conclusions are different from
those drawn from other theories which suggest an infinite number of
magic doping fractions in the superconductors \cite{Chen2004,Feng2006}.

\begin{figure}[b]

\begin{centering}
\resizebox{1\columnwidth}{!}{ \includegraphics{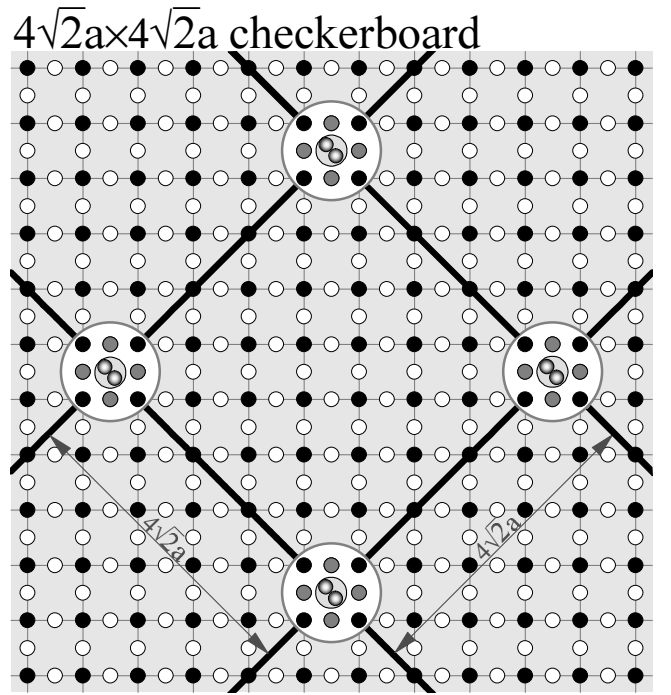}}
\par\end{centering}

\caption{The $4\sqrt{2}a\times4\sqrt{2}a$ nondispersive checkerboard in hole-doped
$CuO_{2}$ plane.\textbf{ }For 3D superconductors, this checkerboard
pattern is experimentally unobservable, as discussed in the text.}

\label{fig4}
\end{figure}

\begin{figure}
\begin{centering}
\resizebox{1\columnwidth}{!}{ \includegraphics{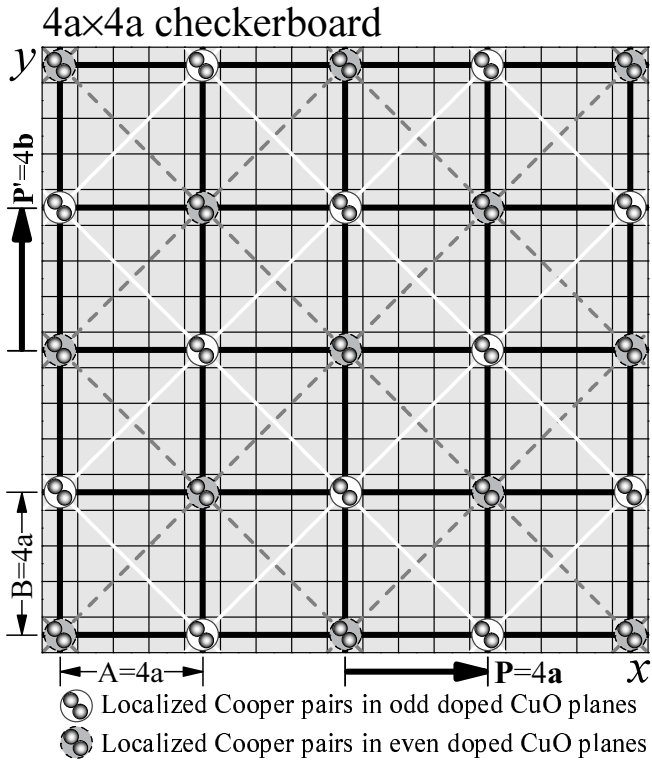}}
\par\end{centering}

\caption{The tetragonal phase with $4a\times4a$ checkerboard in hole-doped
cuprates at doping level $x=$1/16.\textbf{ }The doped $CuO_{2}$
planes can be divided into `odd doped $CuO$ planes' and `even doped
$CuO$ planes' and there is a displacement $\mathbf{P}=4\mathbf{a}$
(or $\mathbf{P'}=4\mathbf{b}$) between them, as indicated in the
figure. This implies that though a single $CuO_{2}$ exhibits the
\textbf{$4\sqrt{2}a\times4\sqrt{2}a$} checkerboard order (white or
gray dashed lines), the results of STM experiment on this sample show
still a \textbf{$4a\times4a$} pattern (black lines).}

\label{fig5}
\end{figure}

For the tetragonal phases of $x=$1/18, 2/25, 1/8 and 2/9, the \textbf{$6a\times6a$,
$5a\times5a$,} $4a\times4a$ and \textbf{$3a\times3a$} checkerboard
patterns can be easily determined from Fig. \ref{fig2}a and equation
(\ref{x_NSC1}), respectively. While for the octahedral phases of
$x=$1/16, 1/9 and 1/4, the experimental results of checkerboard structures
must be $4a\times4a$, $3a\times3a$ and $2a\times2a$, respectively,
which are different from those proposed by Fig. \ref{fig2}c. This
is nothing to be surprised about as the checkerboard structures of
the individual $CuO_{2}$ can still be described by Fig. \ref{fig2}c.
To illustrate this more explicitly, let us consider the case of $x=$1/16
which has been experimentally confirmed to show the $4a\times4a$
checkerboard in the LSCO sample \cite{Zhou2004}. According to our
theory, the localized Cooper pairs in a single doped $CuO_{2}$ plane
of cuprates exhibit the $4\sqrt{2}a\times4\sqrt{2}a$ checkerboard
pattern at $x=$1/16, as shown in Fig. \ref{fig4}. In the octahedral
phase, the doped $CuO_{2}$ planes of the cuprate may be divided into
`odd doped CuO plane' and `even doped CuO planes' and there is a displacement
between them $\mathbf{P}=4\mathbf{a}$ (or $\mathbf{P'}=4\mathbf{b}$),
as indicated in Fig. \ref{fig5}. It can be seen clearly from the
figure that though a single $CuO_{2}$ shows the $4\sqrt{2}a\times4\sqrt{2}a$
pattern at $x=$1/16, the results of STM experiment on this sample
may still reveal the global $4a\times4a$ structure \cite{Zhou2004}.

Our results support the experimental observation that the suppression
of $T_{c}$ and the existence of $4a\times4a$ order in cuprates around
$x=$1/8 \cite{Hanaguri2004} and 1/16 \cite{Zhou2004}. And the abnormal
suppression of the superconductivity at $x=$1/9 \cite{Wu2009} and
1/18 \cite{Keimer1992,Kastner1998} has also been confirmed by many
researchers. In 2005, Komiya et al. \cite{Komiya2005} reported hole-doping
dependence of the in-plane resistivity in $LSCO$ samples, they find
a tendency towards charge ordering at $0.06(\sim1/16)$, $0.09(\sim3/32)$,
$0.13(\sim1/8)$, and $0.18(\sim3/16)$. One may find that the results
of their experiments are rough to some extent, more sophisticated
experiments may reveal that the peaks of 0.06 and 0.09 are in fact
double degenerate. According to our theory, the 0.06 peak corresponds
to two adjacent magic doping fractions of $1/18\approx0.0556$ and
$1/16\approx0.0625$, while the 0.09 peak may be contributed by $2/25\approx0.080$
and $1/9\approx0.111$. Furthermore, in our framework, the number
of 3/16 is not the magic number in cuprates. Let us pay attention
to the experimental results of Fig. 2b in Komiya et al. paper \cite{Komiya2005},
which clearly show a stronger peak around $x=0.22$ (apparently different
from the suggested $x=3/16=0.1875$) consistent with our prediction
of $x=2/9\approx0.222$. For the largest magic number $x=1/4=0.25$,
the corresponding non-superconducting $2a\times2a$ checkerboard phase
is more unstable because of a much stronger pair-pair interaction
inside the superconductor. Consequently, the suppression of $T_{c}$
at $x=$1/4 can only be observed at a much lower temperature.

It is worth noting that all the available experimental data of the\textbf{
}`magic numbers' are included in our theory. However, for exactly
the same question, two researcher groups \cite{Chen2004,Feng2006}
have derived two totally different expressions indicating `infinite
magic numbers' in the $p$-type superconductors, while at the same
time some typically `magic numbers' (such as 1/9 and 1/18) have been
excluded from their expressions. Hence, it is reasonable to argue
that they may also be on the wrong track.

\section{Rotational symmetry breaking }

Broken symmetries have been detected in many hole-doped high-temperature
superconductors when they undergo a phase transition \cite{Daou2010}.
The nature of the broken symmetry in the non-superconducting pseudogap
phase is a central problem in the effort to understand the pseudogap
in the high-$T_{c}$ copper oxide superconductors. Recently, it has
been experimentally confirmed that the breaking of 90\textdegree{}-rotational
symmetry may occur within every $CuO_{2}$ unit cell in underdoped
$Bi_{2}Sr_{2}CaCu_{2}O_{8}$, which can be regarded as the best support
for the model of Fig. \ref{fig1}.

\begin{figure}
\begin{centering}
\resizebox{1\columnwidth}{!}{ \includegraphics{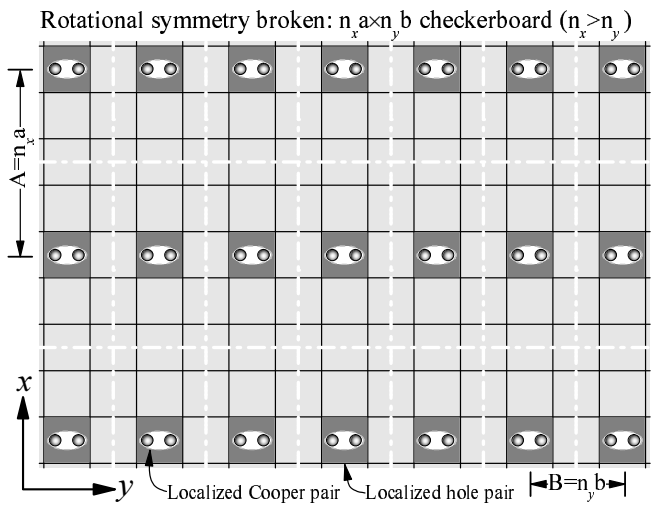}}
\par\end{centering}

\caption{The checkerboard pattern with a rotational symmetry broken. The localized
Cooper pairs (or localized hole pairs) self-organize into some quasi-one-dimensional
Peierls chains. }

\label{fig6}
\end{figure}

The localized hole pairs (see Fig. \ref{fig1}) are the intra-unit-cell
states of the pseudogap phase with an intrinsically broken of the
four-fold rotational symmetry within single $CuO_{2}$ unit cell.
This apparent broken symmetry may be the key to understand the pseudogap
phase of copper-oxide superconductors \cite{Huang3}. Similar to the
liquid crystals, the significant ultra short-range anisotropic orientational
structure of Fig. \ref{fig1} may eventually lead to the long-range
orientational order in the superconductors, as shown in Fig. \ref{fig6}.
Since we assume $n_{x}>n_{y}$, the macroscopic four-fold rotational
symmetry of the checkerboard patterns of Fig. \ref{fig2} are broken.
In addition, it is not difficult to find from Fig. \ref{fig6} that
the polarized Cooper pairs (or localized hole pairs) can self-organize
into some minimum-energy quasi-one-dimensional Peierls chains. For
the special case of $n_{y}=1$, the localized Cooper pairs form some
periodic orders (quasi-two-dimensional vortex lattices) of the most
compact Peierls chains which we considered as the superconducting
ground states \cite{Huang1}. By applying an external field on the
superconductor along $y$-direction, there is a charge ordering phase
transition from the most compact Peierls chains (the superconducting
ground state) to the periodic chains (the superconducting excited
state) with the minimum electron-electron distance $\delta=b/2$.

\section{Conclusions}

We have shown for the first time the intrinsic relation between the
magic doping fractions and checkerboard patterns in hole-doped cuprates.
It has been proved theoretically that there exist merely seven magic
doping fractions numbers ($x=$1/18, 1/16, 2/25, 1/9, 1/8, 2/9 and
1/4) in the superconductors, which are completely different from those
of quantum field theory, suggesting the existence of an infinite magic
doping fractions in the systems. In our view, the phenomenon of the
completely destruction of superconductivity is a macro-reflection
of the localization of all the micro paired electrons in the superconductor.
Physically, the pinning of a huge amount of electrons must always
accompanied by the appearance of a high symmetry localized electronic
state which is characterized by a definite and periodic checkerboard
pattern. For the seven magic doped samples, all of the checkerboard
structures have been uniquely determined in this study. Our theoretical
framework supports the new finding of the symmetries broken within
one copper-oxide unit in the superconductors. We are confident that
a `new window' to the mysterious world of superconductivity has been
opened.

\end{document}